\documentclass[]{jfm}

\usepackage{graphicx}
\usepackage{newtxtext}
\usepackage{newtxmath}
\usepackage{natbib}
\usepackage{hyperref}
\hypersetup{
    colorlinks = true,
    urlcolor   = blue,
    citecolor  = black,
}

\newcommand{\RomanNumeralCaps}[1]
\linenumbers

\title[Shock wave oscillations over axisymmetric bodies in high-speed flow]{Large- and small-amplitude shock wave oscillations over axisymmetric bodies in high-speed flow}

\author[V. Sasidharan and S. Duvvuri]{Vaisakh Sasidharan\ns\and\ns Subrahmanyam Duvvuri\ls\corresp{\email{subrahmanyam@iisc.ac.in}}}

\affiliation{Turbulent Shear Flow Physics and Engineering Laboratory\\Department of Aerospace Engineering, Indian Institute of Science, Bengaluru 560 012, India}

\begin{document}
\maketitle

\begin{abstract}
The phenomena of self-sustained shock wave oscillations over conical bodies with a blunt axisymmetric base subject to uniform high-speed flow are investigated in a hypersonic wind tunnel at Mach number $M = 6$. The flow and shock wave dynamics are dictated by two non-dimensional geometric parameters presented by the three length scales of the body, two of which are associated with the conical forebody and one with the base. Time-resolved schlieren imagery from these experiments reveals the presence of two disparate states of shock wave oscillations in the flow, and allows for the mapping of unsteadiness boundaries in the two-parameter space. Physical mechanisms are proposed to explain the oscillations and the transitions of the shock wave system from steady to oscillatory states. In comparison to the canonical single-parameter problem of shock wave oscillations over spiked-blunt bodies reported in literature, the two-parameter nature of the present problem introduces distinct elements to the flow dynamics.
\end{abstract}

\begin{keywords}
high-speed flow, shock waves.
\end{keywords}

\vspace*{-1.5 cm}
\section{Introduction}
\label{sec:intro}
The presence of shock waves in some compressible flow scenarios can lead to flow unsteadiness. A commonly encountered example of this is the unsteadiness generated from interactions between a shock wave and boundary layer flow, where an adverse pressure gradient imposed by the flow geometry or the shock wave leads to boundary layer separation. The separation bubble, described by a separation length scale $L_{sep}$, generates a separation shock that exhibits unsteady oscillatory motion along a region of length $L_{i}$ upstream of the bubble. Such interactions typically occur in high-speed flow deflection over a ramp/fin/protuberance or with impingement of an oblique shock onto a surface (see figure 1); here the ratio $L_{i}/L_{sep}$ is observed to be approximately 0.3 \citep{Dussauge06}. These types of flows can broadly be classified under a category of small-amplitude shock wave oscillations. In contrast, large-amplitude shock wave oscillations is the key characteristic associated with terminal shock unsteadiness over a transonic airfoil \citep{Lee01} and buzz instability in high-speed air intakes \citep{Seddon99}.

\begin{figure}
\begin{center}
\includegraphics[trim=2.3cm 2.35cm 2.35cm 1.6cm,clip,width=0.99\textwidth]{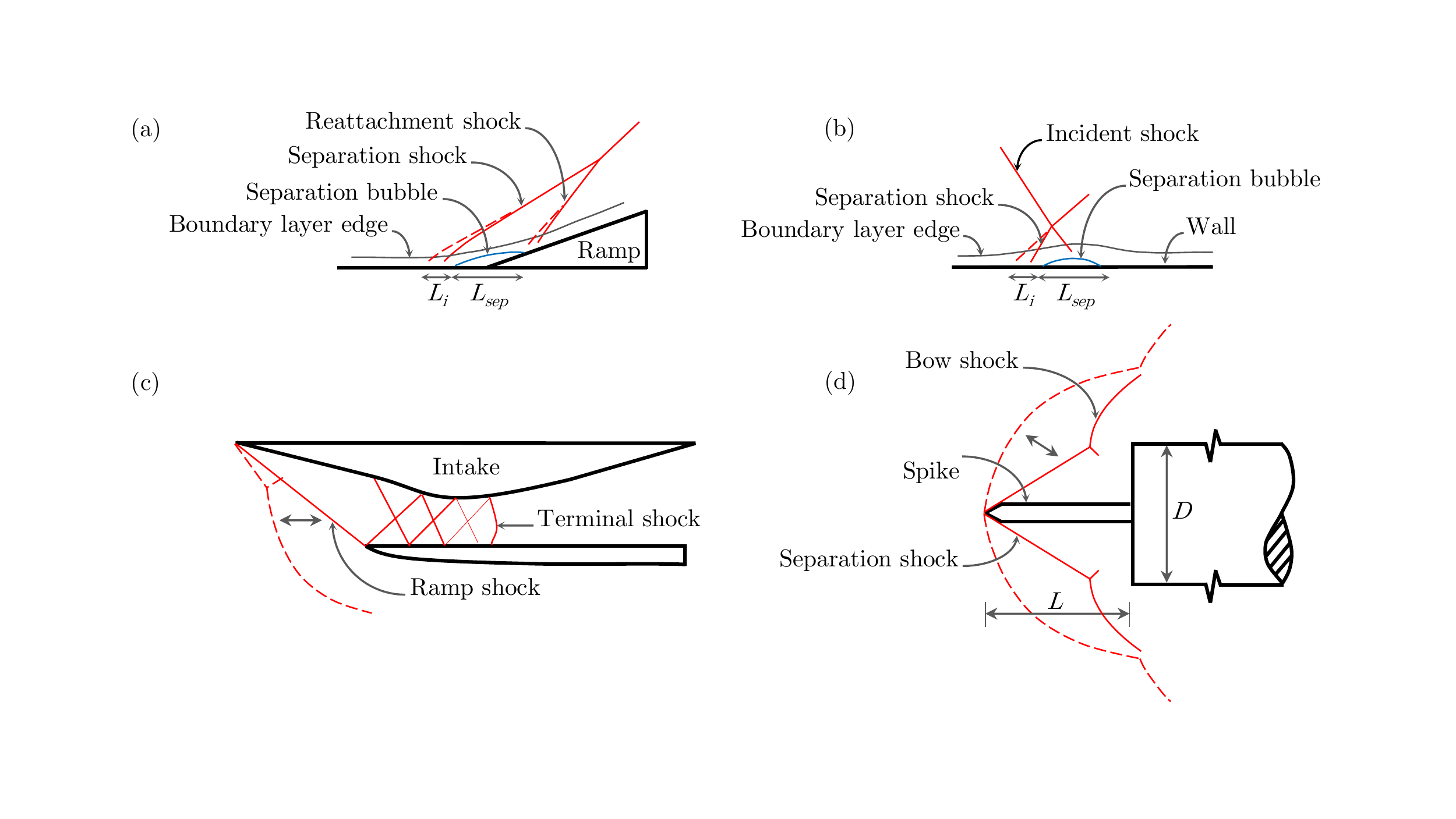}
\caption{A schematic illustration of some flows that exhibit shock wave oscillations. Top and bottom panels show examples with small- and large-amplitude oscillations respectively.}
\vspace*{-2 mm}
\end{center}
\end{figure}

The canonical and seemingly simple geometry of an axisymmetric spiked cylinder, shown in figure 1d, also exhibits large-amplitude shock wave oscillations in high-speed flow for combinations of spike length $L$ and base cylinder diameter $D$ that approximately fall in the range $0.2 < L/D < 1.5$ \citep[see][and references therein]{Kenworthy78,Panaras81,Feszty04a,Panaras09}. These large-amplitude oscillations, termed as \emph{pulsations} in literature, are self-sustained and are characterized by unsteady and periodic shock wave motion and separated flow along almost the entire spike length. Interestingly, an increase in $L/D$ into the range $1.5 < L/D < 2.5$ results in distinct small-amplitude shock wave oscillations, termed simply as \emph{oscillations}, characterized by periodic flipping of the leading separation shock wave between convex and concave shapes \citep{Kenworthy78,Feszty04b}. A steady shock wave system forms around the body outside of these ranges, \emph{i.e.} for $L/D < 0.2$ and $L/D > 2.5$ (the values for $L/D$ boundaries cited here are based on experimental evidence and are found to slightly vary with flow Mach number). The important aspect to note for this class of spiked-blunt body problems is that the dynamics are governed by a single non-dimensional geometric parameter, \emph {i.e.} $L/D$.

The present work aims to understand shock wave oscillations in a class of problems with two geometric parameters. With the exception of a very recent computational study of the canonical double cone problem by Hornung et al. (under review), large-amplitude shock wave oscillation problems governed by more than a single geometric parameter have not been subject to detailed study. The geometry chosen for this exercise is a forward-facing circular cylinder with a right circular conical forebody (see figure 2a). This geometry can be fully described by three independent length scales -- base cylinder diameter $D$, forebody cone length $L$, and cone base diameter $d$ -- which naturally gives two independent non-dimensional parameters, taken here to be $L/d$ and $D/d$ (note that the cone half-angle $\theta\,=\,\tan^{-1}[d/2L]$). Flow behavior across the two-parameter space was studied by extensive wind tunnel experimentation at Mach number $M = 6$. Both large- and small-amplitude shock wave oscillations were observed in the experiments, and physical mechanisms are proposed to explain the observations. The following sections presents experimental details followed by results and discussions.

\begin{figure}
\begin{center}
\includegraphics[trim=1.4cm 2.85cm 1.15cm 1.6cm,clip,width=1\textwidth]{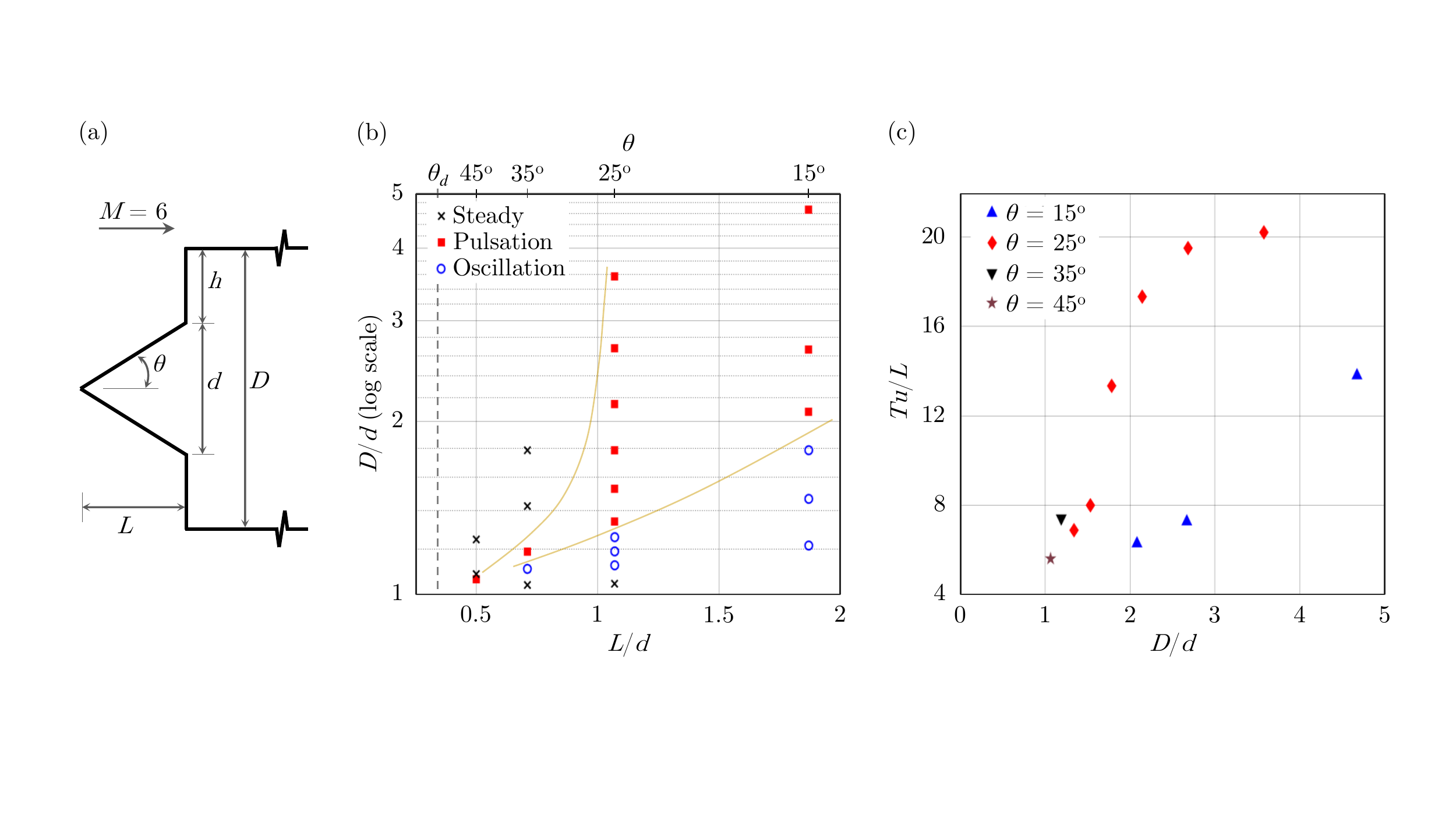}
\caption{(a) Axisymmetric model geometry; (b) $L/d$-$D/d$ parameter space with classification of shock wave behavior at experimental data points into three states. The solid-line curves represent empirical boundaries between the states; (c) Non-dimensional pulsation time period $Tu/L$.}
\vspace*{-2 mm}
\end{center}
\end{figure}

\section{Experimental results and discussion}
\label{sec:exp}
\subsection{Hypersonic wind tunnel experiments}
\label{sec:exp}
Experiments were performed in the Roddam Narasimha Hypersonic Wind Tunnel (RNHWT) at IISc. RNHWT is 0.5 meter diameter enclosed free-jet facility that can be operated in the Mach number range 6 to 10. All the present experiments were carried out at $M=6$ and a free-stream unit Reynolds number of $7\times10^6$ m$^{-1}$, with a corresponding free-stream velocity $u = 900$ m/s. Test models with various combinations of $D$, $L$, and $d$ were used; for reference dimensions of the largest model used are $D$ = 100 mm, $L$ = 40 mm, and $d$ = 80 mm. Four different values of $L/d = [0.5,\, 0.71,\, 1.07,\, 1.87]$, corresponding to $\theta = [45^{\circ},\, 35^{\circ},\, 25^{\circ},\, 15^{\circ}]$, were studied, each at different $D/d$ values in the range $1 < D/d < 5$. Note that the conical shock wave detachment angle at Mach 6 is $\theta_{d} = 55.4^{\circ}$. Shock wave location and motion were visualized by employing the schlieren technique in a time-resolved manner by using a high-power pulsed diode laser (Cavilux Smart, 640 nm wavelength, 10 ns pulse width) as the light source and a high-speed camera (Phantom V1612) for imaging. The short pulse width of the light source allows for a high degree of spatial localization in imaging fast-moving shock waves. Data was recorded in the range of 48,000 to 160,000 frames-per-second and provides good temporal resolution for detailed analysis of flow features. Figure 2b provides an overview of all the combinations of $L/d$ and $D/d$ studied here; data markers in the figure denote individual experiments. Based on the shock wave system behavior, each location (or data marker) in the $L/d$-$D/d$ parameter space is classified into one of three states: \emph{steady}, \emph{pulsations}, and \emph{oscillations}. The terminology used here is borrowed from spiked cylinder literature referenced above, where pulsations and oscillations refer to large- and small-amplitude shock wave oscillations respectively. The boundary curves between states shown in figure 2b are empirically inferred based on the available data points and are qualitative in nature. Observations and discussions for the three flow states and transitions between them are presented next. Note that the terminology of \emph{transition} in the present context refers to a change in the flow from one state to another, brought about by changes to the governing geometric parameters. Hereon shock waves will be referred in short as \emph{shocks}. 

\subsection{Steady shock wave systems}
\label{sec steady}

\begin{figure}
\begin{center}
\includegraphics[trim=1.55cm 3.7cm 2.15cm 2.3cm,clip,width=1\textwidth]{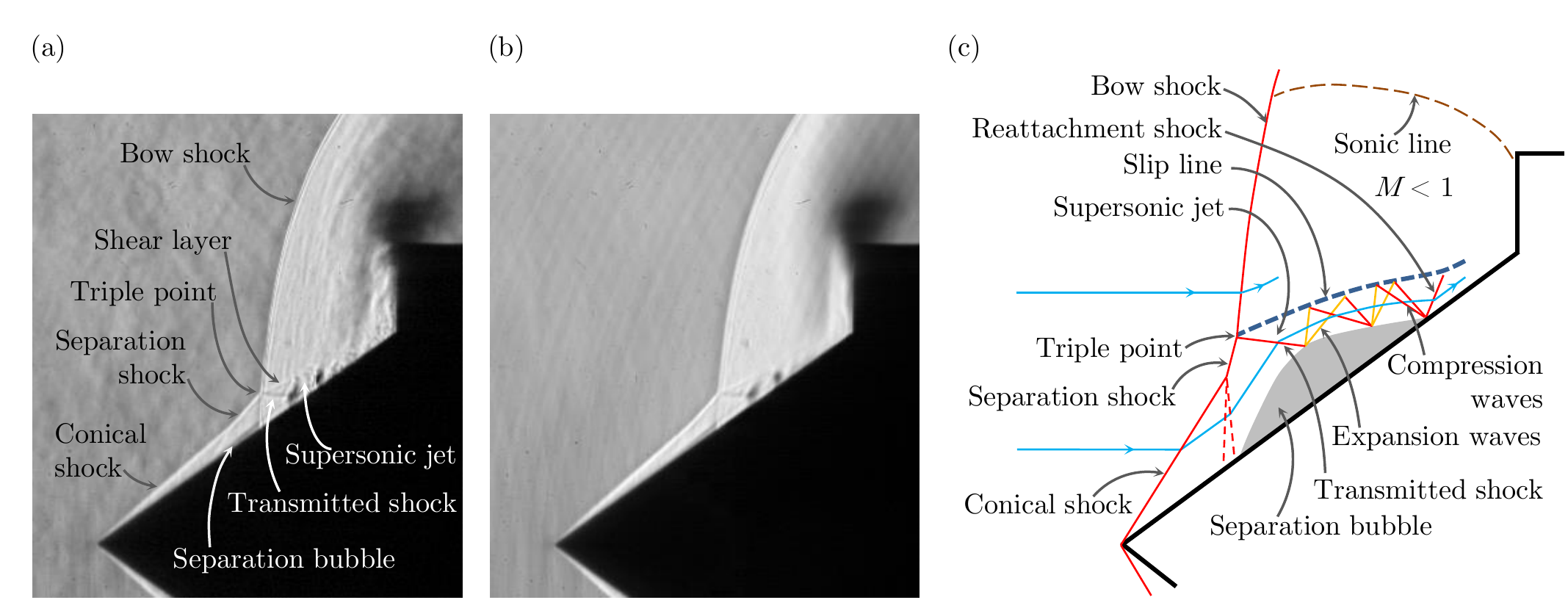}
\caption{(a) An instantaneous schlieren image of the steady shock system for $\theta = 35^{\circ}$ and $D/d = 1.43$ (also see supplementary video file); (b) An average intensity map obtained from a temporal sequence of 5000 schlieren images for the same $\theta$ and $D/d$; (c) A schematic illustration (not to scale) of key flow features of the steady shock system.}
\vspace*{-2 mm}
\end{center}
\end{figure}

For any given $L/d$ it is easy to understand the formation of a steady shock system for trivial cases at limiting values of $D/d$, \emph{i.e.} a large value of $D/d$ (where $D/d$ is sufficiently greater than $L/d$) and a value of $D/d$ very close to 1. In the former case the base cylinder gives rise to a steady leading bow shock with the conical forebody situated entirely in the downstream subsonic region. In the latter case a steady conical shock forms around the forebody (for $\theta < \theta_d$) along with a downstream expansion region centered at the base cylinder shoulder.

Now consider the case of $L/d = 0.71$ ($\theta = 35^{\circ}$) for which formation of non-trivial steady shock systems were observed for $D/d \gtrapprox 1.3$ (see figure 2b). As $D/d$ is reduced starting from a large value, the bow shock stand-off distance reduces and eventually the shock moves downstream of the cone tip and leads to scenarios where an attached conical shock forms on the forebody. The schlieren images in figure 3 show an example of this for $D/d = 1.43$ $\theta = 35^{\circ}$, and the key features of the shock structure are schematically illustrated in figure 3c (given the axisymmetric nature of the flow, only the top half of the flow field is shown in all the schlieren images presented here). The shock structure seen in the schlieren images can be understood in the following sequential order. The initial transmitted shock generated at the intersection of the conical and bow shocks impinges on the boundary layer flow over the cone surface. The adverse pressure jump across this shock results in partial separation of the boundary layer and formation of a separation bubble, which in turn generates a separation shock with a shock angle larger than the conical shock (angles with reference to the cone axis). It is this separation shock that then intersects the bow shock and leads to the formation of the final transmitted shock and the steady shock system seen in figure 3. The intersection point of separation, bow, and transmitted shocks is referred to as the \emph{triple point}. The nature of interaction between the separation and bow shocks falls under the type IV classification of \citet{Edney68}. The difference in flow velocities downstream of the bow and transmitted shocks results in the development of a shear layer with subsonic and supersonic flow above and below the velocity slip line respectively. Downstream of the transmitted shock a supersonic shock train forms in region between the shear layer and the wall, typical of type IV interactions, and this is referred to as a \emph{supersonic jet}. It is to be noted that the terminology of \emph{shock train} is used here in a general sense to describe a series of reflected compression and expansion waves, and/or shocks. The extent of this shock train is dependent on the transmitted shock Mach number (typically, higher the Mach number, longer is the extent) and the downstream pressure condition. The subsonic flow downstream of the shock train curves upwards, goes through the sonic line, and accelerates to supersonic velocities around the base cylinder shoulder. Note that the part of the incoming free-stream flow that passes through the conical shock and the supersonic shock train that follows experiences a lower total pressure loss in comparison to the flow that passes through the bow shock (see the two streamlines illustrated in figure 3c) as dictated by fundamental gas dynamics. For $\theta = 35^{\circ}$ a distinct steady state at $D/d$ close to 1 was also observed; this represents a trivial case as per above discussion and does not contain flow features of any interest. 

Formation of non-trivial steady shock systems with the above features were also observed at $L/d = 0.5$ ($\theta = 45^{\circ}$) for $D/d > 1.07$. However, an important distinction is noted -- the separation bubble was found to be relatively smaller in size and therefore does not lead to interaction between the separation and bow shocks, while other features like formation of a shear layer and a shock train remain the same. This distinction can be attributed to the relatively large cone half-angle which results in a stronger conical shock and lower downstream Mach number, and thereby a weaker transmitted shock that leads to a more localized separation bubble. For $L/d = 1.07$ ($\theta = 25^{\circ}$) a trivial steady state was observed at $D/d$ close to 1, similar to $\theta = 35^{\circ}$. Interestingly, non-trivial steady states were not observed for $\theta = 25^{\circ}$ at larger values of $D/d$ that are below the $D/d$ threshold beyond which the bow shock moves upstream of the cone tip; $D/d = 3.57$, the largest value in this study for $\theta = 25^{\circ}$, was experimentally found to be very close to the threshold value. It therefore is concluded that $\theta = 25^{\circ}$ cannot support a steady shock system (excluding the trivial scenarios). Similarly, a steady state was not observed for $L/d = 1.87$ ($\theta = 15^{\circ}$), with $D/d = 4.57$ being the largest value investigated. It is to be noted that as $\theta$ is further reduced toward the limit $\theta = 0^{\circ}$, the geometry approaches a spiked cylinder, where a steady state was observed in literature for $D/L \gtrapprox 5$ \citep{Kenworthy78}; this is a trivial case with the bow shock standing upstream of the spike tip. This implies that in the scenario of $\theta \to 0^{\circ}$ ($L \gg d$), only trivial steady shock systems are observed at very large values of $D/d$ where $D/L \gtrapprox 5$. Therefore by qualitative interpolation of the steady-pulsation boundary between $\theta = 25^{\circ}$ and $\theta \to 0^{\circ}$ it is hypothesized that non-trivial steady states do not exist for $\theta$ below a critical value, which lies somewhere between $35^{\circ}$ and $25^{\circ}$.

\subsection{Shock wave pulsations}
\label{sec:pulsations}
\subsubsection{Transition between steady and pulsation states}
\begin{figure}
\begin{center}
\includegraphics[trim=6.3cm 0.2cm 6.3cm 0cm,clip,width=0.99\textwidth]{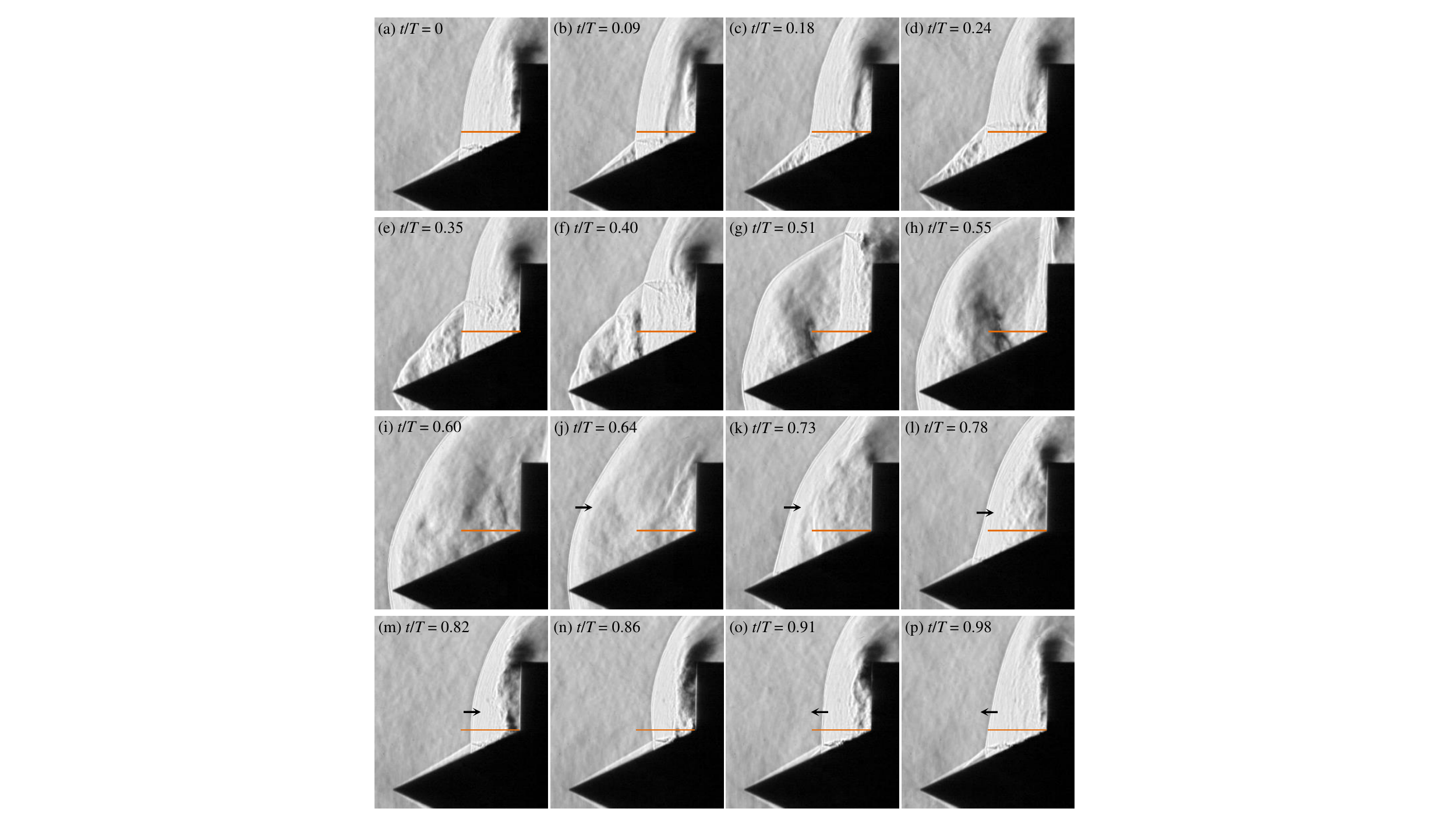}
\caption{A selection of schlieren images in sequence at sixteen different instances over one pulsation time period $T$ for $D/d = 2.14$ and $\theta = 25^{\circ}$. A body-fixed reference line (in color) is provided in all images as a visual aid to track shock wave motion (also see supplementary video file).}
\vspace*{-3 mm}
\end{center}
\end{figure}

Data from these experiments exhibit a clear trend by which transition from non-trivial steady to pulsation state occurs with decrease in $D/d$ for fixed $L/d$ and for increase in $L/d$ for fixed $D/d$. A physical mechanism to explain this behavior is proposed by examining the influence of two parameters: base cylinder shoulder height $h = (D-d)/2$ and strength of the attached conical shock. Consider the steady case of $L/d = 0.71$ ($\theta = 35^{\circ}$) and $D/d = 1.43$ shown in figure 3. Starting with this steady flow, a decrease in $D/d$ for fixed $L/d$ brings about a reduction in the non-dimensional shoulder height $h/d$. With that the bow shock retracts closer to the base cylinder, and thereby the triple point also moves closer to the base cylinder and further away from the cone axis. The transmitted shock then impinges on the cone surface (hereon referred to as \emph{wall}) at a location closer to the cone base. Note that the pressure jumps along the shock train and the inward turning of the wall at the cone base, by an angle $(90-\theta)^{\circ}$, result in static pressure increase along the wall downstream of the triple point. The average magnitude of this adverse pressure gradient is set by the stagnation pressure at the cone base and the distance along the wall over which the pressure rise occurs. Now with a decrease in $D/d$, the total mass flow through the conical shock increases given the increase in distance between the triple point and cone axis, and for the same reason the mass flow through the bow shock reduces. This reduction in the mass flow through the bow shock is in addition to the reduction brought about by the decrease in $h/d$ which shortens the vertical extent of the shock. Therefore a relatively larger mass of fluid with higher total pressure (in comparison to the total pressure downstream of the bow shock) is introduced by the shock train in the base region. Hence the stagnation pressure at the cone base plausibly increases, and certainly does not decrease, with a decrease in $D/d$. This leads to the conclusion that a decrease in $D/d$ brings about an increase in the gradient of adverse wall pressure, chiefly due to the change in the transmitted shock impingement location. As $D/d$ is reduced below some critical value, the increasing gradient is expected to induce significant flow reversal near the wall, and allow for mass influx into the separation bubble and therefore a growth in its size. As will be seen in section 2.3.2, unsteady growth of the separation bubble drives shock pulsations, and hence characterizes the onset of pulsations. From figure 2b it is observed that the critical $D/d$ value for this transition lies between 1.43 to 1.19 for $\theta = 35^{\circ}$.

Now consider the reverse transition, \emph{i.e.} from pulsation to steady state, for decrease in $L/d$ (increase in $\theta$) at fixed $D/d$. For instance at $D/d = 1.53$ the shock system undergoes such a transition going from $\theta = 25^{\circ}$ to $35^{\circ}$. In this scenario, $h/d$, and therefore the bow shock stand-off distance remains nearly unchanged, whereas the cone half-angle increases, thereby increasing the conical shock angle and shock strength. The stronger shock results in relatively higher total pressure loss, thereby reducing the wall pressure gradient downstream of the triple point. Further, the reduction in the inward turning angle at the cone base also alleviates to some extent the adverse pressure gradient along the wall. Above a critical value of $\theta$ (which falls in between $\theta = 25^{\circ}$ and $\theta = 35^{\circ}$ for $D/d = 1.53$) these effects curtail flow reversal and arrest the growth of the separation bubble, and thereby restore stability to the shock wave system. The above discussions can be generalized for other values of $D/d$ and $L/d$ where transitions between steady and pulsation states are observed.

\subsubsection{Pulsation mechanism}
During pulsations the shock structure is severely disrupted from its steady state, and the shock system executes periodic motion with a time period $T$ and amplitudes comparable to the length scale $L$ of the conical forebody. As a representative example, figure 4 shows a selection of schlieren images in sequence at sixteen different instances over one pulsation cycle for $D/d = 2.14$ and $\theta = 25^{\circ}$. The starting point for the cycle, \emph{i.e.} time $t/T = 0$ (figure 4a), is chosen to be the instant where the bow shock stand-off distance and its shape above the triple point match the bow shock that forms over a forward-facing cylinder of the same diameter without the conical forebody (which was obtained by a separate experiment). At $t/T = 0$ the instantaneous shock wave and flow structure show a close resemblance to a steady state, like seen in figure 3a. Growth of the separation region, for reasons outlined in section~2.3.1, results in upstream motion of the separation point along with a continuous increase in the separation shock angle, and consequently upward motion of the triple point; this is seen in figures 4a through 4d. The separation region continues to grow and distorts the shape of the separation shock, which now begins to resemble a bow shock as seen in figures 4e and 4f. During this phase the bow shock stand-off distance remains unchanged. Following this the triple point rapidly moves downstream as it continues its upward motion, and the separation shock transforms itself into a bow shock, as seen in figures 4f through 4i. Figure 4i shows the instance where this transformation is complete and the newly-formed bow shock fully envelopes the axisymmetric body. At this point in the cycle the sonic region around the cylinder shoulder (denoted in figure 3c by a sonic line) reaches its maximum area and allows for a high mass flux. The increased mass flux around the cylinder shoulder results in a rapid size reduction of the separated region and thereby a pull back of the newly-formed bow shock (figures 4j through 4n). The conical shock re-appears during this pull back phase and interacts with the retreating bow shock. Note that bow shock overshoots the reference stand-off distance ($t/T = 0$) during pull back and reaches closer to the base; this is clearly seen in figure 4n which shows the instant where the retreating bow shock comes to a stop before reversing its direction of motion. Following the direction reversal, the bow shock returns to the stand-off position held at beginning of the cycle (figures 4n through 4p) and the next cycle of pulsation begins.

The key features of shock pulsation are growth of the separation region, deformation of the shock structure and upward motion of the triple point, and rapid collapse of the separation region and deformed shock structure. These features were consistently observed in the schlieren data across the entire region that exhibits pulsations in the $L/d$-$D/d$ parameter space. However, changes in relative values of $L/D$ and $D/d$ within the pulsation region bring about some differences in the deformed shock structures. Particularly for low values of $D/d$, close to the transition from pulsations to oscillations, there is no formation of a large bow shock like the one observed in figure 4. The non-dimensional time periods $Tu/L$ for all the pulsation cases shown in figure 2b are reported in figure 2c. It is interesting to note the steady-pulsation transition boundary trend seen in figure 2b indicates that with increasing $\theta$ the shock unsteadiness either completely ceases beyond some value of $\theta$ slightly larger than $45^\circ$, or is confined to an increasingly small region of $D/d$ close to 1 until $\theta = \theta_d$. This observation can be attributed to reduction in the transmitted shock strength with increasing $\theta$ (as discussed in section 2.2), which implies that beyond a certain $\theta$ the wall pressure gradient is perhaps not sufficiently strong to induce flow reversal and trigger large-scale unsteadiness.

\subsection{Shock wave oscillations}
\label{sec:oscillations}
\begin{figure}
\begin{center}
\includegraphics[trim=1.1cm 4.4cm 1cm 3.4cm,clip,width=1\textwidth]{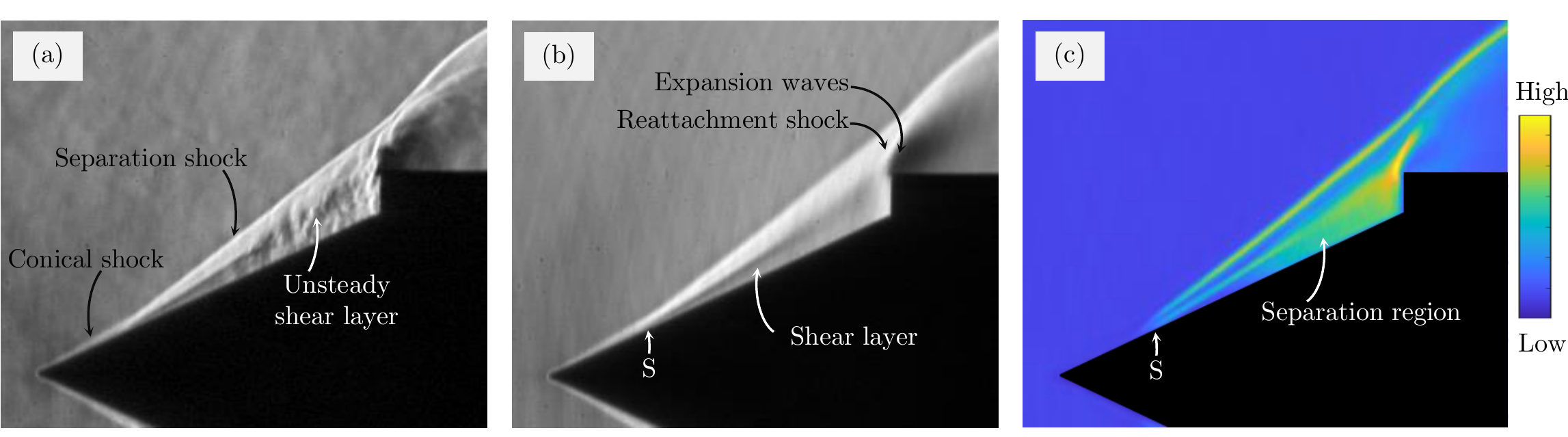}
\caption{(a) An instantaneous schlieren image of the oscillating shock system for $\theta = 25^{\circ}$ and $D/d = 1.26$ (also see supplementary video file); (b) An average (temporal) intensity map; (c) A standard deviation (temporal) intensity map. `S' denotes the separation point.}
\vspace*{-2 mm}
\end{center}
\end{figure}

Consider the pulsation region of the parameter space for reducing values of $D/d$ at $L/d = 1.07$ ($\theta = 25^{\circ}$) . At $D/d \approx 1.3$, the nature of shock oscillations undergoes a distinct change from large-amplitude unsteadiness to small-amplitude fluctuations in shock structure. A mechanism for this transition is proposed by considering the flow behavior with increasing $D/d$ starting at the limiting value of $D/d = 1$, \emph{i.e.} $h/d = 0$. As $h/d$ is gradually increased, a separation bubble begins to grow in the vicinity of the cone base, nested by the corner. This situation is similar to the canonical problem of high-speed flow over a forward-facing step \citep[\emph{e.g.}][]{Murugan16}, albeit here the step has a downstream inclination. The separation bubble will be accompanied by the development of a shear layer in the region over the bubble, driven by the velocity gradient between the subsonic flow inside the bubble and supersonic flow outside. For small bubble sizes, \emph{i.e.} for relatively small values of $h/d$, this shear layer remains steady, and overall the shock system is steady. The steady states recorded for $\theta = 25^\circ$ and $35^\circ$ at $D/d$ close to 1 (see figure 2b) are representative of this scenario. With increasing $h/d$, the bubble size increases and pushes the separation shock upstream along the conical surface, and leads to a longer development length of the shear layer. Instabilities in the shear layer begin to manifest and grow beyond a critical value of the development length; this is clearly observed in figure 5a which shows an instantaneous schlieren image for the case $D/d = 1.26$ and $\theta = 25^{\circ}$ as a representative example for shock oscillations. These instabilities interact with the separation shock resulting in small-amplitude high-frequency undulations in its structure. Further, these shear layer instabilities are accompanied by small-amplitude expansions and contractions in the size of the separation region due to impingement of the unsteady shear layer on the cylinder base, and this results in periodic fore-and-aft motion of the separation point along the conical surface. The motion of the separation point naturally imparts unsteadiness to the separation shock. These flow features are collectively termed as \emph{oscillations} for the purposes of classification and making a distinction from pulsations. Figures 5b and 5c show the average and standard deviation respectively of image intensity obtained from a temporal sequence of 5000 schlieren images for $D/d = 1.26$ and $\theta = 25^{\circ}$. The separation shock clearly stands out in the standard deviation map as a region of relatively large fluctuations in intensity, indicative of the unsteadiness caused due to its interaction with unsteady shear layer structures along with the unsteadiness brought about by the fore-and-aft motion of the separation point. The unsteadiness within the separation region is also highlighted by the standard deviation map.

The size of the separation bubble ceases to scale with $D/d$ beyond a certain threshold, and a bow shock forms around the base cylinder and interacts with the separation shock. This brings into play the pulsation mechanisms discussed in section~2.3, and at this point the shock wave system makes a transition from oscillation to a pulsation state; this transition is observed at $D/d \approx 1.3$ for $\theta = 25^{\circ}$. This understanding can be generalized for other values of $L/d$ ($\theta$) where an increase in $D/d$ starting in the oscillation regions brings about a transition in the shock system from oscillation to pulsation state, with the $D/d$ transition boundary dependent on $L/d$. The pulsation-oscillation boundary trend in the $L/d-D/d$ parameter space also indicates a narrowing of the unsteadiness regime in $D/d$ with increasing $\theta$, and lends support to the conclusions drawn in section 2.3.2 on the basis of the steady-pulsation boundary behavior.

\section{Conclusions}
Shock behavior over a conical body with a blunt axisymmetric base is explained by this detailed experimental study. A steady and two distinct oscillatory states of the shock system were identified, and the boundaries between these states were mapped out in the governing two-parameter space. Interplay between the physical effects brought about by the base cylinder shoulder height (in terms of the bow shock stand-off distance) and the cone half-angle (in terms of the conical shock strength) determine the nature of shock unsteadiness and transition boundaries. Large-amplitude shock oscillations (pulsations) are driven by periodic unsteady growth and collapse of the separated flow region that forms over the conical surface. Whereas the small-amplitude shock oscillations are primarily driven by instabilities in the shear layer that forms over a corner separation bubble. Experimental evidence indicates that below a critical value of $\theta$, which lies somewhere in between $35^{\circ}$ and $25^{\circ}$ at $M =6$, the flow cannot sustain a non-trivial steady shock system. Further, shock unsteadiness either completely ceases beyond an other critical value of $\theta$ that lies somewhere in between $45^{\circ}$ and $\theta_d$ at $M =6$, or is restricted to a narrow region in $D/d$ for $45^{\circ}<\theta<\theta_{d}$. Given that the bow shock stand-off distance and the conical shock strength depend on the free-stream Mach number $M$, the transition boundaries given by this study will undergo a shift with changes in $M$. However, the qualitative features of shock unsteadiness and driving mechanisms are expected to broadly remain the same.

\section*{Supplementary material}
The digital video file ``S1.mp4'' contains schlieren videos from experiments corresponding to the instantaneous schlieren images shown in figures 3, 4, and 5.



\bibliographystyle{jfm}


\end{document}